\documentclass[12pt,preprint]{aastex}
\def\gtorder{\mathrel{\raise.3ex\hbox{$>$}\mkern-14mu
                \lower0.6ex\hbox{$\sim$}}}
\def\ltorder{\mathrel{\raise.3ex\hbox{$<$}\mkern-14mu
                \lower0.6ex\hbox{$\sim$}}}

\shorttitle{CN and HCN in PDRs}
\shortauthors{Boger \& Sternberg}

\begin{document}
\title{Bistability in Interstellar Gas-Phase Chemistry}
\vspace{1cm}
\author{Gai I. Boger and Amiel Sternberg}
\vspace{0.5cm}
\affil{School of Physics and Astronomy and the Wise Observatory,
        The Beverly and Raymond Sackler Faculty of Exact Sciences,
        Tel Aviv University, Ramat Aviv 69978, Israel}
\email{amiel@wise.tau.ac.il}

\begin{abstract}
We present an analysis of ``bistability''
in gas-phase chemical models of dark interstellar clouds.
We identify the chemical mechanisms that
allow high- and low-ionization solutions to the
chemical rate-equations to coexist.
We derive simple analytic scaling relations
for the gas densities and ionization rates
for which the chemistry becomes bistable. 
We explain why bistability is sensitive to 
the H$_3^+$ dissociative recombination rate coefficient, 
and why it is damped by gas-grain neutralization.  

\end{abstract}

\keywords{ISM:molecules -- molecular processes}


\section{Introduction}
\label{introduction}

In this paper we reexamine and analyze the phenomenon of 
``bistability'' that occurs in chemical models of dark interstellar clouds.  
Bistability was first discussed by Le Bourlot et al.~(1993), 
who found that multiple solutions to the chemical rate-equations
sometimes appear in numerical
computations of the gas-phase abundances of atomic and molecular 
species for steady-state dark cloud conditions. When bistability occurs,
three sets, rather than a single unique set, of 
chemical abundances are predicted for identical
input cloud parameters such as
the total gas density, the cosmic-ray (or X-ray)
ionization rate, and the gas-phase supply of heavy elements.
Under such circumstances, the
solution that is converged to in a numerical simulation, and perhaps, the actual
steady-state of a real physical system, depend on the 
initial conditions and history of the gas.

The subject of bistability begins with
Pineau des Forets et al.~(1992) who studied how the
computed chemical states of dark clouds
depend on the gas density and ionization rate.
They found, as would be expected, 
that for a given ionization rate the fractional ionization 
of the gas decreases roughly as the square-root of the gas density.
However, they also found that the ionization fraction and associated
chemical composition changes abruptly, or quasi-discontinuously, 
at a specific critical transition density.
Below the critical density
the clouds are in what Pineau des Forets et al.~referred to as
a ``high-ionization'' phase (HIP), characterized by
high abundances of atomic ions. 
For densities greater than the
critical density 
the clouds are in a ``low-ionization'' phase (LIP), with
high abundances of molecular ions.
Pineau des Forets et al.~also identified a chemical
instability that, they argued, is responsible for the abrupt transition 
that they encountered in their calculations. Bistability was subsequently
discovered by Le Bourlot et al.~(1993) who found that
instead of a phase change at a specific critical density,
the LIP and HIP ``branches'' can sometimes overlap 
for a narrow range of densities.
A third (unstable) branch also appears,
intermediate between the LIP and HIP solutions.

Further investigations have explored the effects of the assumed 
gas phase elemental abundances, ionization rates, gas temperatures, chemical 
reaction rate coefficients, and gas-grain interactions, on the existence 
and character of the multiple solutions (Le Bourlot et al.~1995ab; 
Shalabiea \& Greenberg 1995; Lee et al.~1998; 
Pineau des Forets \& Roueff 2000; 
Viti et al.~2001; Charnley \& Markwick 2003).  These studies have shown 
that bistability is indeed related to the transition from the HIP to LIP
(or vice versa), and that it is  
sensitive to quantities such as the gas-phase heavy-element
abundances (Le Bourlot et al.~1995; Viti et al.~2001), and the adopted
chemical network (Lee et al.~1998), including specifically
the dissociative recombination rate coefficient of H$_3^+$
(Pineau des Forets \& Roueff 2000).
In this paper we explain why the HIP and LIP can overlap,
and we identify the chemical mechanisms that allow the possibility
of multiple solutions.

We will show that bistability 
can be simply understood and analyzed using just a handful of
reactions. In a nutshell, when bistability occurs
atomic ions (usually S$^+$ ions) are the dominant positive charge
carriers for both HIP and LIP conditions. 
Two modes of neutralization are then available to the gas.
In the HIP, neutralization proceeds by slow radiative recombination.
In the LIP, neutralization occurs by rapid dissociative recombination,
via the formation of intermediate molecular ions
in reactions with O$_2$ molecules
(e.g., ${\rm S^+ + O_2 \rightarrow SO^+ + O}$).
Bistability occurs when
a low O$_2$ abundance in the HIP can be
maintained up to maximal gas densities that are {\it greater}
than the minimal densities above which a large O$_2$ density
is maintained in the LIP. We will show by means of a 
simple analysis why this is possible. The explanation involves a
chemical instability, of a type first
introduced by Pineau des Forets et al.~(1992), but not given sufficient
attention in subsequent studies.

Bistability may be of primarily theoretical interest since,
as we will discuss,
it is a purely gas-phase chemical effect.  
If realistic gas-dust neutralization processes are included, 
the multiple solutions are damped or eliminated entirely 
(cf.~Shalabiea \& Greenberg 1995). We will illustrate this
by considering the effect 
of grain-assisted ion-electron recombination
(Weingartner \& Draine 2001) on the chemistry.
Nevertheless, because bistability has been 
the subject of considerable discussion in the literature, we feel that 
a fresh analysis of how, why, and when it occurs is warranted. 

The plan of our paper is as follows. In \S 2 we present
results of an illustrative numerical computation 
in which bistability occurs.
In \S 3 we discuss the key chemical instability
that drives the steady-state solutions to either the
LIP or HIP. In \S 4 we 
derive simple analytic formulae that show how the range
of gas densities and ionization rates
for which the gas becomes bistable depends on the
elemental abundances and several important chemical
rate coefficients. In \S 5 we analyze the sensitivity
of bistability to the H$_3^+$ dissociative recombination
rate coefficient.
In \S 6 we explain why gas-grain neutralization processes
damp and eliminate bistability. We summarize in \S 7.


\section{Computations}

We first illustrate the phenomenon of bistability in a detailed
numerical computation.  For this purpose we
calculate the steady-state atomic and molecular
abundances for dark cloud conditions, as functions of the gas density.
We include only gas-phase formation and destruction
processes, except for molecular hydrogen (H$_2$), which we
assume is formed on grain surfaces. 
The chemistry is driven by cosmic-ray impact ionization
of hydrogen and helium, followed by standard
sequences of ion-molecule, charge-exchange, and neutral-neutral reactions.
The chemistry is also mediated by
radiative and dissociative recombination,
and cosmic-ray induced photoionization and photodissociation.
The effects of externally incident FUV photons are excluded.

We use the Boger \& Sternberg (2005) network
for a system consisting of 70 atomic and molecular
carbon, nitrogen, oxygen, sulfur, and silicon bearing species
(see also Sternberg \& Dalgarno 1995).
We neglect all other heavy elements.
Our set of 868 reactions and rate coefficients 
is based mainly on the {\small UMIST99} compilation
(Le Teuff et al.~2000) with some modifications and alterations
\footnote{Our input list of rate coefficients is available at
ftp://wise3.tau.ac.il/pub/amiel/pdr.}.
As in Boger \& Sternberg (2005) we 
adopt $\zeta$ $Oph$ gas-phase abundances, with
C/H=$1.32\times 10^{-4}$ (Cardelli et al.~1993), 
N/H=$7.50\times 10^{-5}$ (Meyer et al.~1997),
O/H=$2.84\times 10^{-4}$ (Meyer et al.~1998),
Si/H=$1.78\times 10^{-6}$ (Cardelli et al.~1994), and
S/H=$8.30\times 10^{-6}$ (Lepp et al.~1988).
We set He/H=0.1. We assume a constant gas temperature of $T=50$ K
for the entire range of gas densities that we consider.

In our calculations
we solve the equations of chemical equilibrium
\begin{equation}\label{ceqns}
\sum_{jl}k_{ijl}(T)n_jn_l + \sum_j \xi_{ij}n_j
= n_i\biggl\lbrace \sum_{jl}k_{jil}n_l + 
\sum_j\xi_{ji} \biggr\rbrace 
\end{equation}
for the densities $n_i$ of all species $i$ in our set. In these equations,
$k_{ijl}(T)$ are the temperature-dependent 
rate coefficients (cm$^3$ s$^{-1}$) for chemical
reactions between species $j$ and $l$ that lead to the
production of $i$, and $\xi_{ij}$ are the cosmic-ray destruction
rates (s$^{-1}$) of species $j$ with products $i$. The
parameters $\xi_{ij}$ include impact ionizations
of hydrogen and helium by the cosmic-ray protons and
secondary electrons, as well as photo-destruction of heavy
atomic and molecular species by induced UV photons.
The $\xi_{ij}$ are all proportional to
$\zeta\equiv10^{-17}\zeta_{-17}$ s$^{-1}$, the cosmic-ray ionization rate
of molecular hydrogen (Sternberg et al.~1987; Gredel et al.~1989)

The densities $n_i$ must also satisfy the supplementary equations
of element and charge conservation. These are,
\begin{equation}\label{sup1}
\sum_{i} \alpha_{im} n_i = X_m n_{\rm tot}
\end{equation}
where $\alpha_{im}$ is the number of atoms of element $m$
contained in species $i$, $X_{m}$ is the total abundance of
element $m$ relative to the total density
$n_{\rm tot}$ of hydrogen nuclei, and
\begin{equation}\label{sup2}
\sum_{i} q_i n_i = 0
\end{equation}
where $q_i$ is the charge of species $i$. In our computations,
the negative charge is carried entirely by free electrons.

H$_2$ molecules are formed primarily on dust grains
and are destroyed by cosmic-ray ionization. A separate
equation may then be written for the H/H$_2$ balance,
\begin{equation}\label{heqn}
\zeta n_{\rm H_{2}} = Rn_{\rm tot}n_{\rm H}
\end{equation}
where $n_{\rm H}$ and $n_{\rm H_{2}}$ are the atomic
and molecular hydrogen densities respectively, and
$R=3\times 10^{-18}T^{1/2}$ cm${^3}$ s$^{-1}$ is the H$_2$ 
grain surface formation rate coefficient (Biham et al.~2002; 
Vidali et al.~2004).

For fixed gas temperature and gas-phase elemental abundances, 
the density fractions $x_i\equiv n_i/n_{\rm tot}$ as determined
by equations~(\ref{ceqns})-(\ref{heqn}) depend on the single parameter
$n_{\rm tot}/\zeta$ (e.g., Lepp \& Dalgarno 1996; Lee et al.~1998),
as is readily seen by dividing each term
in these equations by $n_{\rm tot}^2$.
We consider a range of parameters for which
almost all of the hydrogen is in the form of H$_2$,
so that $n_{\rm tot}=2n_{\rm H_2}$.
We therefore present our results as functions of the
ratio $n_{\rm H_2}/\zeta_{-17}$ (cm$^{-3}$). 

We solve equations~(\ref{ceqns})-(\ref{heqn}) numerically via 
Newton-Raphson (NR) iteration. In our procedure we vary the
density along the HIP and LIP branches and use the
solutions from the previous density step as the ``initial
guess'' for the NR iteration in the next density step. In the bistability
region we carry out repeated NR iterations with random initial guesses,
so that all three branches are found.

In Figure 1 we plot the densities $n_i/\zeta_{-17}$, and
density fractions $n_i/n_{\rm H_2}$, as functions of 
$n_{\rm H_2}/\zeta_{-17}$
ranging from 10$^2$ to 10$^4$ cm$^{-3}$, for several species
of interest. Panels (a) and (b)
display our results for electrons, C$^+$, S$^+$, 
H$_3^+$, and a sum over all molecular ions. Panels (c) and (d)
display our results for the neutral species C, O, CO and O$_2$.
Several important features are apparent in Figure 1, some of which
have been noted in previous discussions.

For our assumed elemental abundances and gas temperature,
a high-ionization phase (HIP) is present for densities less than 
$(n_{\rm H_2}/\zeta_{-17})_{\rm max}=675$ cm$^{-3}$, 
and a low-ionization phase (LIP) 
is present for densities greater than 
$(n_{\rm H_2}/\zeta_{-17})_{\rm min}=370$ cm$^{-3}$.
The HIP and LIP overlap, and the gas is
bistable, between these minimal and
maximal densities.  The HIP and LIP solutions
are connected by a third branch that we refer to here as the
MIX solution.

In the HIP, the positive charge is carried entirely by atomic ions.
S$^+$ is the dominant ion down to $n_{\rm H_2}/\zeta_{-17}=60$ cm$^{-3}$.
At lower densities C$^+$ takes over, and for $n_{\rm H_2}/\zeta_{-17}<1.1$
cm$^{-3}$, H$^+$ dominates. In the LIP, 
the molecular ions HCO$^+$ and H$_3$O$^+$ are
the primary positive charge carriers
for densities greater than $n_{\rm H_2}/\zeta_{-17}>6\times 10^3$ cm$^{-3}$.
However, S$^+$ dominates at lower densities in the LIP.
Most importantly, in our calculation S$^+$ is the
primary positive charge carrier in {\it both} the HIP and LIP
for the densities at which bistability occurs. 

The fractional ionization increases from a minimum value
$x_{\rm e}=6\times 10^{-6}$ 
as the density decreases from $(n_{\rm H_2}/\zeta_{-17})_{\rm max}$
in the HIP. In the LIP, the fractional ionization decreases from
a maximum value of $x_{\rm e}=2\times 10^{-6}$ 
as the density increases from $(n_{\rm H_2}/\zeta_{-17})_{\rm min}$.
The fractional ionization drops by a factor of $\sim 10$ 
in moving from the HIP to LIP in the bistability region.

The free atomic carbon density and the C/CO abundance 
ratio is large in the HIP.
The ratio ${\rm C/CO}\approx 0.3$ in the bistability region, and
exceeds unity for 
$n_{\rm H_2}/\zeta_{-17} \lesssim 55$ cm$^{-3}$.
In the LIP the C density and the C/CO ratio are small.
In our computation the C/CO ratio drops by a factor
$\sim 50$ in moving from the HIP to the LIP. The
drop in C/CO is associated with an
{\it increase} in the O$_2$ density and the O$_2$/CO ratio.
It is very significant that the
C and O$_2$ densities vary in
opposite directions in the transition between the HIP and LIP.

On the MIX branch the fractional ionization, and the C/CO and
O$_2$/CO ratios, are intermediate between the HIP and LIP.
S$^+$ is also the dominant positive ion on the MIX branch.  
However, the fractional ionization increases with 
gas density along the MIX. 

In Table~1 we list the densities
of selected atomic and molecular species for the three HIP, LIP, and MIX
solutions, for $n_{\rm H_2}=4.3\times 10^2$ cm$^{-3}$ and 
$\zeta=1.0\times 10^{-17}$ s$^{-1}$. We also list the HIP/LIP
density ratios for these values of $n_{\rm H_2}$ and $\zeta$.
Atomic ions, e.g., S$^+$, Si$^+$, and
C$^+$, are abundant in the HIP solution. 
In the LIP solution, molecular ions such as H$_3^+$,
HCO$^+$, and H$_3$O$^+$ are abundant, leading 
to more OH, H$_2$O, and O$_2$ than in the HIP.
The oxygen bearing molecules, NO, SO, and SiO, are enhanced by the
large O$_2$ density that is maintained in the LIP.
Carbon bearing
molecules, such as CN and HCN, are enhanced in the HIP because 
of the large atomic carbon density in this phase
(see also Boger \& Sternberg 2005). CS is produced by reactions
of SO (enhanced in the LIP)
with C (enhanced in the HIP)
so that the CS density changes by only a factor
of $\sim 2$ in moving from the LIP and HIP.


\section{Chemistry and Instability}

Figure 1 shows that as the gas density $n_{\rm H_2}$ is increased along the 
sequence of HIP solutions an abrupt jump to the LIP branch
occurs when $n_{\rm H_2}$ rises above $(n_{\rm H_2}/\zeta_{-17})_{\rm max}$.
Similarly, as $n_{\rm H_2}$ is decreased along the LIP, a jump to the HIP
occurs when $n_{\rm H_2}$ falls below $(n_{\rm H_2}/\zeta_{-17})_{\rm min}$.
As we now discuss, these jumps are signatures of a simple
chemical instability involving electrons, S$^+$ and H$_3^+$ ions,
C atoms, and O$_2$ molecules.
The instability can be understood by considering the
gas-phase formation and destruction sequences
for these species,
for gas densities where the bistable solutions occur.
We first discuss the HIP and LIP, and we then consider the MIX. 

We begin with H$_3^+$. In both the HIP and LIP, this ion
is formed by cosmic-ray ionization
of H$_2$ followed by proton transfer,
$$
{\rm H_2 + cr \rightarrow H_2^+ + e}              \eqno(R1)
$$
$$
{\rm H_2^+ + H_2 \rightarrow H_3^+ + H} \ \ \ .   \eqno(R2)
$$
In the HIP, the electron fraction is large, and
dissociative recombination
$$
{\rm H_3^+ + e \rightarrow H_2 + H}               \eqno(R3)
$$
$$
{\rm H_3^+ + e \rightarrow H + H + H}     \eqno(R4)
$$
dominates the removal of the H$_3^+$.
However, in the LIP the electron fraction is small,
and the H$_3^+$ is removed by
$$
{\rm H_3^+ + CO \rightarrow HCO^+ + H_2} \ \ \ .    \eqno(R5)
$$
For a gas temperature of 50 K, 
dissociative recombination dominates the destruction of
H$_3^+$ when $n({\rm e})/n({\rm CO}) > 10^{-2}$. 

In both the HIP and LIP, the production of O$_2$ is initiated by
proton transfer
$$
{\rm O + H_3^+ \rightarrow OH^+ + H_2} \ \ \ ,        \eqno(R6)
$$
which is followed by abstraction
$$
{\rm OH^+ + H_2 \rightarrow H_2O^+ + H}         \eqno(R7)
$$
$$
{\rm H_2O^+ + H_2 \rightarrow H_3O^+ + H} \ \ \ ,       \eqno(R8)
$$
dissociative recombination
$$
{\rm H_3O^+ + e \rightarrow OH + H_2}             \eqno(R9)
$$
$$
{\rm H_3O^+ + e \rightarrow H_2O  + H} \ \ \ ,    \eqno(R10)
$$
and the neutral-neutral reaction
$$
{\rm OH + O \rightarrow O_2 + H} \ \ \ .          \eqno(R11)
$$
However, the O$_2$ is destroyed
in different ways in the two phases. In the HIP,
the density of free atomic carbon is
large, and the O$_2$ is removed by
$$
{\rm O_2 + C \rightarrow CO + O} \ \ \ .          \eqno(R12)
$$
In the LIP, the C density is small, and
the O$_2$ is removed much less efficiently in reactions
with less abundant atoms and atomic ions. An upper limit
on the O$_2$ density in the LIP is set by
cosmic-ray induced photodissociation
$$
{\rm O_2 + crp \rightarrow O + O} \ \ \ .              \eqno(R13)
$$
Reaction (R13) is inefficient, and the O$_2$
density becomes large in the LIP.

In the HIP, C$^+$ is produced by cosmic-ray driven  
helium impact ionization of CO, 
$$
{\rm He + cr \rightarrow He^+ + e}                \eqno(R14)
$$
$$
{\rm He^+ + CO \rightarrow C^+ + O + He} \ \ \ .          \eqno(R15)
$$
This is then followed by rapid charge transfer
$$
{\rm S + C^+ \rightarrow S^+ + C}        \eqno(R16)
$$
which leads to the production of S$^+$ and C.
The S$^+$ ions are removed by radiative recombination,
$$
{\rm S^+ + e \rightarrow S + \nu} \ \ \ .         \eqno(R17)
$$
The C atoms are removed by cosmic-ray induced photoionization
$$
{\rm C + crp \rightarrow C^+ + e} \ \ \  .        \eqno(R18)
$$
Because cosmic-ray photoionization is inefficient, a large C
density is maintained in the HIP.

In the LIP, C$^+$ is also produced by helium impact ionization
([R14] and [R15]). However, because the O$_2$ density
is large in the LIP, the C$^+$ ions are rapidly neutralized by
the sequence
$$
{\rm C^+ + O_2 \rightarrow CO^+ + O}              \eqno(R19)
$$
$$
{\rm CO^+ + e \rightarrow C + O} \ \ \ ,           \eqno(R20)
$$
and the C$^+$ density becomes small.
In the LIP, charge transfer between C$^+$ and S ([R16])
contributes negligibly to the removal of C$^+$, or to the
formation of S$^+$.
The main source of S$^+$ in the LIP is cosmic-ray induced photoionization
$$
{\rm S + crp \rightarrow S^+ + e} \ \ \ .        \eqno(R21)
$$
Crucially, O$_2$ also dominates the neutralization of S$^+$ 
in the LIP, via
$$
{\rm S^+ + O_2 \rightarrow SO^+ + O}              \eqno(R22)  
$$
$$
{\rm SO^+ + e \rightarrow S + O} \ \ \ .            \eqno(R23)  
$$
For a gas temperature of 50 K,
neutralization of S$^+$ by O$_2$ dominates
over radiative recombination
when $n({\rm O}_2)/n({\rm e}) > 0.8$.
Finally, the C atoms formed by (R20) are also removed
by reactions with O$_2$ ([R12]), and the C density becomes small in the LIP.

For the gas densities of interest, most of the gas-phase sulfur
is present as S and S$^+$ in both the HIP and LIP, 
with only a small fraction bound in
molecules. S$^+$ is the primary 
positive charge carrier in both the HIP and LIP. 
However, in the HIP S$^+$ is removed by
radiative recombination, whereas in the LIP by reactions with O$_2$.
In the HIP,
H$_3^+$ is removed by dissociative recombination, 
C is removed by cosmic-ray induced
photoionization, and O$_2$ is removed by C. In the LIP, H$_3^+$ is removed by
CO, C is removed by O$_2$, and O$_2$ is limited by 
cosmic-ray induced photodissociation. The abundance ratio C/O$_2$ is
large in the HIP, and is small in the LIP.

Now consider a gas where S$^+$ is again the dominant charge carrier,
but which differs from the HIP or LIP in that dissociative recombination
contributes significantly to the removal of H$_3^+$ (as in the HIP),
while on the other hand S$^+$ is neutralized by O$_2$ (as in the LIP).
Assume also that O$_2$ is produced by the sequence (R6)-(R11),
and that the O$_2$ destruction rate
increases if the O$_2$ density decreases.
This is possible when reaction (R12) between O$_2$ and C 
is the dominant destruction mechanism for
both species, so that $n({\rm C})$ increases if $n({\rm O_2})$ decreases.
For these conditions a chemical instability can occur, as we
demonstrate with the following perturbation analysis.

Consider a possible solution to the steady-state rate
equations. With the above assumptions this requires
$n({\rm e})=n({\rm S^+})$. 
Now consider a perturbative 
increase in the ionization fraction $x_{\rm e}\equiv n({\rm e})/n_{\rm H_2}$,
carried out at fixed gas density $n_{\rm H_2}$.
This will reduce H$_3^+$ by a factor that depends on the
relative efficiencies of dissociative recombination 
versus reaction with CO in removing the H$_3^+$.
The O$_2$ is removed by C (via [R12]), so for a given C density
the reduction in H$_3^+$ will lead
to an equal reduction in O$_2$ (since O$_2$ is formed
by [R6]-[R11])
\footnote{In this analysis we assume that most of the
oxygen not locked in CO remains atomic.}.
However, the decrease in O$_2$ will lead to an increase 
in C, since the C is removed by O$_2$ (again via [R12]). This reduces
O$_2$ further still. The S$^+$ density then increases by a factor
equal to the total reduction in O$_2$ (since S$^+$ is removed by O$_2$). 
Finally, to close the loop, $n({\rm e})$ must then be set equal
to the new value of $n({\rm S^+})$, since by assumption S$^+$ carries
the positive charge. However, if the increase
in S$^+$ is larger than the initial increase in the
electron density the cycle ``runs away'' in a positive
feedback loop.  

This H$_3^+$-O$_2$-S$^+$ cycle is illustrated
schematically in Figure 2.  It is clear that the cycle
can also lead to a run-away decrease in the electron density,
in response to a perturbative decrease in $n({\rm e})$.
For increasing electron densities, O$_2$ decreases and
C increases. For decreasing electron densities,
O$_2$ increases and C decreases.

The instability shuts off for sufficiently large or small fractional ionizations.
For large $x_{\rm e}$, radiative recombination,
rather than removal by O$_2$, becomes the dominant 
S$^+$ neutralization process. 
For small $x_{\rm e}$, reaction with CO replaces 
dissociative recombination as the
H$_3^+$ removal mechanism. Either way, the H$_3^+$-O$_2$-S$^+$ 
instability cycle shuts off, 
and the gas becomes chemically stable. 

In the HIP, dissociative recombination dominates
the destruction of H$_3^+$, but S$^+$ is removed by
radiative recombination. In the LIP, S$^+$ is
removed by O$_2$, but H$_3^+$ is removed by CO.
Crucially therefore, the H$_3^+$--O$_2$--S$^+$ cycle is 
inoperative on the HIP and LIP branches,
and these solutions are chemically stable.

We now identify the maximum possible density for the HIP,
$(n_{\rm H_2}/\zeta_{-17})_{\rm max}$, as the
point at which O$_2$, {\it as formed and destroyed in the HIP}, becomes
sufficiently abundant to begin
removing the S$^+$.  When this occurs,
the H$_3^+$--O$_2$--S$^+$ cycle turns on (since H$_3^+$
is still being removed by electrons)
and a stable HIP is no longer possible. As the density is increased
above $(n_{\rm H_2}/\zeta_{-17})_{\rm max}$
the instability induces a sharp drop in $n({\rm e})$.
As the electron density drops in the run-away loop, 
O$_2$ increases and C decreases. The system is finally
stabilized when the electron density becomes sufficiently small
for reactions with CO to begin dominating the removal of H$_3^+$.
The H$_3^+$--O$_2$--S$^+$ cycle then turns off, and a stable solution on the LIP
is reached.

Similarly, we identify the minimum possible density for the LIP,
$(n_{\rm H_2}/\zeta_{-17})_{\rm min}$, as the point at
which the fractional ionization, {\it as determined for the LIP}, becomes
sufficiently large for dissociative recombination to
begin dominating the removal of H$_3^+$.  When this occurs,
the H$_3^+$--O$_2$--S$^+$ cycle turns on, 
(since S$^+$ is still removed by O$_2$)
and a stable LIP is no longer possible. As the density is decreased
below $(n_{\rm H_2}/\zeta_{-17})_{\rm min}$
the instability induces a sharp rise in $n({\rm e})$.
As the electron density increases,
O$_2$ decreases and C increases.
The electron density rises until it becomes sufficiently
large (and the O$_2$ density sufficiently small) for
radiative recombination to dominate the neutralization 
of S$^+$. The instability then
shuts off, and a stable solution on the HIP is reached.

The MIX branch can now also be understood. It represents
the locus of equilibrium solutions, with varying $n_{\rm H_2}$, for which the
H$_3^+$-O$_2$-S$^+$ cycle {\it is} operative, but for which
any change in electron density is precisely matched by a corresponding
change in the S$^+$ density.
For these solutions, dissociative recombination and reactions with CO
both contribute to the removal of H$_3^+$, and S$^+$ is
removed by O$_2$. Furthermore, reaction (R12) is the dominant
destruction mechanism, for
both O$_2$ and C. As the electron fraction increases along the MIX, the 
O$_2$ formation rate decreases, but by a factor
that is smaller than the increase in electron density.
The density of C atoms (which remove the O$_2$) increases
such that the total increase in O$_2$, and the resulting
increase in S$^+$, exactly matches the increase in 
electron density.

Figure 3 summarizes the key characteristics of the
HIP, LIP, and MIX branches, as described above.
The triangular symbols indicate the three
``legs'' of the H$_3^+$-O$_2$-S$^+$ cycle
illustrated in Figure 2.
In the HIP, removal of S$^+$ by O$_2$
is inoperative so this leg is labeled as ``OFF''.
In the LIP, dissociative recombination of H$_3^+$
is inoperative so this leg is ``OFF'' for this phase.
For the MIX, which shares properties of
both the HIP and LIP, all three legs of the H$_3^+$-O$_2$-S$^+$ cycle
operate, and are ``ON''. In Figure 3 we also indicate the
dominant removal reactions for S$^+$, H$_3^+$, O$_2$, and C,
for each of the three branches.

The HIP and LIP
as originally defined by Pineau des Forets et al.~(1992),
were distinguished by whether
H$_3^+$ is removed by dissociative recombination (HIP),
or by reactions with CO (LIP)
\footnote{These authors also recognized that the transition
from the LIP to HIP is driven by a chemical instability 
involving O$_2$. However, they did not consider how the
instability is quenched in each phase.}. 
Here we have identified a second, and crucial, distinction.
In the HIP, atomic ions, 
are removed by (slow) radiative recombination.
In the LIP the atomic ions
are effectively removed by rapid dissociative
recombination, via the formation
of intermediate molecular ions, as in the sequence (R22) and (R23).    

In our discussion it is
S$^+$ that plays the crucial role of positive charge carrier
in both the HIP and LIP.
The reason why simultaneous HIP, LIP, and 
MIX solutions are possible now emerges. Bistability occurs when radiative
recombination is able to dominate the removal of 
S$^+$ in the HIP up to maximal gas densities that are 
{\it greater} than the minimal densities down to which
reactions with CO dominate the removal of H$_3^+$ in the LIP. 
Alternatively, bistability occurs when it is possible for a low
HIP O$_2$ abundance to be maintained up to maximal gas densities
that are greater than the minimal gas densities down to which a large LIP O$_2$
density can be maintained. This is our key result.
When this occurs, there exists a
range of densities for which the S$^+$ ions are either
removed slowly via radiative recombination (HIP, low O$_2$)
or rapidly via dissociative recombination (LIP, high O$_2$).
The simultaneous solutions are separated
by the H$_3^+$-O$_2$-S$^+$ instability cycle, which drives
the O$_2$ density to either small (HIP) or large (LIP) values.

The critical densities
$(n_{\rm H_2}/\zeta_{-17})_{\rm min}$ and
$(n_{\rm H_2}/\zeta_{-17})_{\rm max}$, as defined above, depend on 
a variety of factors including the elemental abundances,
and several specific chemical rate coefficients.
In \S4 we derive simple
analytic expressions for these critical densities.
We determine, analytically, when it is possible for
$(n_{\rm H_2}/\zeta_{-17})_{\rm max}$ to exceed 
$(n_{\rm H_2}/\zeta_{-17})_{\rm min}$,
which is our condition for bistability, and we determine
the range of gas densities for which bistability occurs.


\section{Analysis}
\subsection{HIP and ${\bf n_{\rm max}}$}

For stable HIP conditions the electron density must be sufficiently large for
dissociative recombination to
dominate the removal of H$_3^+$, and the O$_2$ density must be
sufficiently small for radiative recombination
to dominate the removal of S$^+$.
We now show analytically that consistent solutions
that satisfy these HIP conditions are possible provided the
density is less than a critical {\it maximum} gas density.

We first derive an expression for the
O$_2$ density in the low density limit of the HIP.
We show that in this limit the O$_2$ density is small,
and radiative recombination dominates the removal 
of S$^+$.  We then show that our low density expression for
O$_2$ remains consistent only up to a maximal gas density.
We identify this as the maximum possible density for HIP solutions.
We then verify that
dissociative recombination dominates the removal
of H$_3^+$, a necessary condition for the HIP,
for all densities below this critical density.

In the HIP, O$_2$ is produced by the sequence (R6)-(R11),
and is removed by reaction (R12) with atomic carbon. This gives,
\begin{equation}\label{oxy1}
n({\rm O_2}) = \frac{n({\rm H_3^+})n({\rm O})k_6d_{\rm OH}}
                                {n({\rm C})k_{12}}
\end{equation}
where $k_6=8.0\times 10^{-10}$ 
and $k_{12}=3.3\times 10^{-11}$ cm$^3$ s$^{-1}$
are the rate coefficients (at 50 K) of (R6) and (R12),
and $d_{\rm OH}=0.74$ is the fraction (Jensen et al.~2000) of
H$_3$O$^+$ recombinations that produce OH.  
Atomic carbon is
produced by helium impact ionization of CO followed by immediate
charge transfer neutralization ([R15] and [R16]), and
is removed by cosmic-ray induced photoionization ([R18]).
Because (R15) dominates the removal of He$^+$,
the rate of C formation
equals the cosmic-ray ionization rate of helium. Therefore,
\begin{equation}\label{c1}
\frac{n({\rm C})}{n_{\rm H_2}} = \frac{X_{\rm He}}{p_{18}} = 10^{-4} \ \ \ ,
\end{equation}
independent of $n_{\rm H_2}/\zeta$. 
In this expression
$p_{18}=1.02\times 10^3$ is the efficiency of (R18), 
and $X_{\rm He}=0.1$ is the
helium abundance.  It is clear that a large C density is indeed
expected in the HIP, since $n({\rm C})/n_{\rm H_2}=10^{-4}$ is
comparable to the typical total gas-phase carbon abundance
(e.g.~$X_{\rm C}=1.32\times 10^{-4}$ in our numerical model).
The large C density keeps the O$_2$ abundance small in the HIP.

The O$_2$ density is proportional to H$_3^+$,
which in turn, depends on the electron density.
For electrons we set
$n({\rm e})=n({\rm S^+})$.  The S$^+$ is produced
the same way as atomic carbon, i.e., 
by helium impact ionization of CO followed by charge transfer
([R15] and [R16]). We assume that the
S$^+$ is removed by radiative recombination.
This gives,
\begin{equation}\label{efrac1}
x_{\rm e}\equiv
\frac{n({\rm e})}{n_{\rm H_2}}=
\biggl(\frac{X_{\rm He}\zeta}{k_{17}n_{\rm H_2}}\biggr)^{1/2}
=2.9\times 10^{-4}\biggl(\frac{\zeta_{-17}}{n_{\rm H_2}}\biggr)^{1/2}
 \ \ \ .
\end{equation}
where the radiative recombination rate coefficient 
$k_{17}=1.2\times 10^{-11}$ cm$^3$ s$^{-1}$.
In this limit the fractional ionization varies as $(n_{\rm H_2}/\zeta)^{-1/2}$.
For $n_{\rm H_2}/\zeta_{-17} = 100$ cm$^{-3}$, equation~(\ref{efrac1}) gives
$x_{\rm e}=2.9\times 10^{-5}$ in good agreement with $2.3\times 10^{-5}$
computed numerically (see Figure 1).

H$_3^+$ is produced by cosmic-ray ionization and is removed
by dissociative recombination, so that
\begin{equation}\label{h3plus1}
\frac{n({\rm H_3^+})}{n_{\rm H_2}} = 
\frac{\zeta}{k_{3,4} n({\rm e})} = 
\frac{1}{k_{3,4}}
\biggl(\frac{k_{17}\zeta}{X_{\rm He}n_{\rm H_2}}\biggr)^{1/2}
=2.0\times 10^{-7}\biggl(\frac{\zeta_{-17}}{n_{\rm H_2}}\biggr)^{1/2}
\end{equation} 
where $k_{3,4}=1.7\times 10^{-7}$ cm$^3$ s$^{-1}$ 
is the total rate coefficient for dissociative recombination. 
The H$_3^+$ abundance also varies as $(n_{\rm H_2}/\zeta)^{-1/2}$.
For $n_{\rm H_2}/\zeta_{-17} = 100$ cm$^{-3}$, equation~(\ref{h3plus1}) gives
$n({\rm H_3^+})/n_{\rm H_2}=2.0\times 10^{-8}$, as also computed numerically.

Inserting equations~(\ref{c1}), (\ref{efrac1}), and (\ref{h3plus1})
into equation~(\ref{oxy1}) gives
\begin{equation}\label{oxy2}
\frac{n({\rm O_2})}{n_{\rm H_2}} = 
               \frac{2X_{\rm O}^\prime}{X_{\rm He}^{3/2}}
               \frac{k_{17}^{1/2}}{k_{12}}
               \frac{k_6}{k_{3,4}}
               d_{\rm OH}p_{18}
     \biggl(\frac{\zeta}{n_{\rm H_2}}\biggr)^{1/2}
=1.1\times 10^{-5}\biggl(\frac{\zeta_{-17}}{n_{\rm H_2}}\biggr)^{1/2}
\end{equation}
where $X_{\rm O}^\prime\equiv X_{\rm O}-X_{\rm C}=1.5\times 10^{-4}$, and where
we have assumed that most of the oxygen not bound in CO
remains atomic. The O$_2$ abundance varies as $(n_{\rm H_2}/\zeta)^{-1/2}$.
For $n_{\rm H_2}/\zeta_{-17} = 100$ cm$^{-3}$, equation~(\ref{oxy2}) gives
$n({\rm O_2})/n_{\rm H_2}=1.1\times 10^{-6}$, close to
$3.0\times 10^{-7}$ computed numerically.

The O$_2$ density as given by equation~(\ref{oxy2}),
is sufficiently small for 
removal of S$^+$ by O$_2$ to be  
negligible compared to radiative recombination.
This can be seen by considering the ratio of the 
S$^+$ removal rates
\begin{equation}\label{ratSp}
\frac{k_{22}n({\rm O_2})}{k_{17}n({\rm e})} = 
\frac{2X_{\rm O}^\prime}{X_{\rm He}^2}
\frac{k_6k_{22}}{k_{3,4}k_{12}}
d_{\rm OH}
p_{18} = 0.05
\end{equation}
where $k_{22}=1.5\times 10^{-11}$ cm$^3$ s$^{-1}$ is the rate coefficient for
reaction (R22). Because this ratio is
independent of $n_{\rm H_2}/\zeta$, the 
removal of S$^+$ by O$_2$ is negligible at all densities so long
as equation~(\ref{oxy2}) for the O$_2$ abundance is an
appropriate approximation.

Equation~(\ref{oxy2}) is valid 
so long as cosmic-ray photons dominate the removal of
the carbon atoms, so that the C density
is given by equation~(\ref{c1}). However,
cosmic-ray induced destruction occurs at a rate
independent of $n_{\rm H_2}$, whereas the O$_2$ density
as given by equation~(\ref{oxy2}) varies as 
$n_{\rm H_2}^{1/2}$. Therefore, for sufficiently large $n_{\rm H_2}$
reactions with O$_2$ ([R12])
must eventually dominate the removal of the C atoms.
When this occurs
the C density becomes small, and the O$_2$ density
must become large (see equation~[\ref{oxy1}]).
We identify this as
the point where O$_2$ becomes sufficiently abundant to
begin removing the S$^+$. However, because H$_3^+$ continues
to be removed by dissociative recombination,
the H$_3^+$-O$_2$-S$^+$ instability cycle turns on at this point,
and a stable HIP is no longer possible.

Thus, the HIP can be maintained so long as
the O$_2$ density given by equation~(\ref{oxy2})
is small enough for cosmic-ray photoionization 
to dominate the removal of the C atoms. This condition,
\begin{equation}\label{Hcond}
\frac{k_{12}n({\rm O_2})}{p_{18}\zeta} = 
\frac{2X_{\rm O}^\prime}{X_{\rm He}^{3/2}}
\frac{k_6}{k_{3,4}}
k_{17}^{1/2}d_{\rm OH} \biggl(\frac{n_{H_2}}{\zeta}\biggr)^{1/2} 
< 1 
\end{equation}
provides a maximal density for HIP solutions.
For our standard elemental abundances and rate-coefficients
\begin{equation}\label{Hcond2}
\biggl(\frac{n_{\rm H_2}}{\zeta_{-17}}\biggr)_{\rm max} = 7.63\times10^2
\ \ \ {\rm cm}^{-3}
\end{equation}
This is slightly larger than our numerically computed 
density of 675 cm$^{-3}$ 
at which the transition to the LIP occurs (see Figure 1).

For consistency we must verify that dissociative recombination 
dominates the removal of H$_3^+$ for densities as high as
$(n_{\rm H_2}/\zeta)_{\rm max}$.
Dissociative recombination dominates when the ratio
\begin{equation}\label{rat2}
\frac{k_{3,4}n({\rm e})}{k_5n({\rm CO})} = 
\biggl(\frac{X_{\rm He}}{k_{17}}\biggr)^{1/2}
\frac{k_{3,4}}{k_5}
\frac{1}{X_{\rm C}} 
\biggl(\frac{\zeta}{n_{\rm H_2}}\biggr)^{1/2}
\gg 1 \ \ \ ,
\end{equation} 
where $k_5=1.7\times 10^{-9}$ cm$^3$ s$^{-1}$ is the
rate coefficient of reaction (R5).
For $(n/\zeta_{-17})_{\rm max}=7.63\times 10^2$ cm$^{-3}$
this ratio equals $8.0$, so consistency is obtained.

\subsection{LIP and ${\bf n_{\rm min}}$}

In the LIP, the electron density must be sufficiently small for
CO to dominate the removal of H$_3^+$, and the O$_2$ density
must be sufficiently large for 
O$_2$ to dominate the removal of S$^+$.
We now show that these conditions can be satisfied,
provided the gas density is greater than a critical 
{\it minimum} density.

To show this, we first derive an expression for the
electron density assuming that S$^+$ is removed
by abundant O$_2$, with an upper limit for
the O$_2$ density set by cosmic-ray induced photodissociation.
We show that CO then dominates the removal of H$_3^+$
for densities greater than a minimum critical density.
We then verify that the assumption that O$_2$ dominates
the removal of S$^+$ is consistent for densities
greater than this critical density.

For the electron density in the LIP we again assume that 
$n({\rm e})=n({\rm S^+})$. In the LIP,
S$^+$ is produced by cosmic-ray induced photoionization,
and we assume that it is removed by O$_2$. This gives,
\begin{equation}\label{Le1}
\frac{n({\rm e})}{n_{\rm H_2}} = 
\frac{2X_{\rm S}p_{21}\zeta}{k_{22}n({\rm O_2})} 
\end{equation}
where $p_{21}=2.0\times 10^3$ is the efficiency of (R21).
H$_3^+$ is removed by CO, so that
\begin{equation}\label{Le2}
\frac{n({\rm H_3^+})}{n_{\rm H_2}} = 
\frac{\zeta}{k_5n({\rm CO})} = 
\frac{1}{2k_5 X_{\rm C}} 
\frac{\zeta}{n_{\rm H_2}} =
2.2\times 10^{-5}\frac{\zeta_{-17}}{n_{\rm H_2}}
\ \ \ ,
\end{equation}
where we have
assumed that in the LIP all of the carbon is locked in CO.
For $n_{\rm H_2}/\zeta_{-17}=10^3$, equation~(\ref{Le1}) gives
$n_{\rm H_3^+}/n_{\rm H_2}=2.2\times 10^{-8}$, in good agreement
with the numerically computed $1.0\times 10^{-8}$.

For O$_2$ we again assume the production sequence
(R6)-(R11). If the O$_2$ is
removed by only cosmic-ray induced photodissociation,
\begin{equation}\label{Le3}
\frac{n({\rm O_2})}{n_{\rm H_2}} = 
\frac{n({\rm H_3^+})n({\rm O})k_6d_{\rm OH}}{\zeta p_{13}n_{\rm H_2}} =
\frac{k_6}{k_5} 
\frac{1}{p_{13}}
d_{\rm OH} \frac{X_{\rm O}^\prime}{X_{\rm C}}=
2.3\times 10^{-4} \ \ \ ,
\end{equation}
where $p_{13}=1.7\times 10^3$ is the efficiency of (R13).
This is actually an upper-limit for the O$_2$ abundance because
additional chemical reactions contribute to the removal of the
O$_2$ in the LIP. In this approximation the O$_2$ density is
independent of $n_{\rm H_2}/\zeta$, consistent with our numerical
computations (see Figure 1). Because the upper limit is close to
the total available oxygen not locked in CO, we
set $n({\rm O_2})/n_{\rm H_2} \equiv 2f_{\rm O_2}X_{\rm O}^\prime$,
where the O$_2$ fraction $f_{\rm O_2}\lesssim 1$.
In our numerical computation $f_{\rm O_2}\approx0.2$
(see Figure 1).

The electron fraction, $x_{\rm e}$, is then
\begin{equation}\label{Le4}
\frac{n({\rm e})}{n_{\rm H_2}} = 
\frac{X_{\rm S}}{X_{\rm O}^\prime}
\frac{p_{21}}{k_{22}}
\frac{1}{f_{\rm O_2}}
\frac{\zeta}{n_{\rm H_2}} =
7.4\times 10^{-5}\frac{1}{f_{\rm O_2}}\frac{\zeta_{-17}}{n_{\rm H_2}}
\ \ \ .
\end{equation}
In this limit $x_{\rm e}$ varies as $(n_{\rm H_2}/\zeta)^{-1}$.
For $f_{\rm O_2}=0.2$, equation~(\ref{Le4}) gives
$x_{\rm e}=3.7\times 10^{-7}$, in good agreement with
$4.1\times 10^{-7}$ in Figure 1.

The electron density as given by 
equation~(\ref{Le4}) is independent of $n_{\rm H_2}$, whereas the
CO density decreases with $n_{\rm H_2}$.
Reactions with CO can therefore dominate the removal of H$_3^+$ 
only down to a minimal density set by the condition
\begin{equation}\label{Le5}
\frac{k_{5}n({\rm CO})}{k_{3,4}n({\rm e})} = 
\frac{2X_{\rm C}X^\prime_{\rm O}}{X_{\rm S}}
\frac{k_5k_{22}}{k_{3,4}p_{21}}
f_{\rm O_2}
\frac{n_{\rm H_2}}{\zeta}
 > 1  \ \ \ .
\end{equation}
This relation provides a minimal critical density 
\begin{equation}\label{Le6}
\biggl(\frac{n_{\rm H_2}}{\zeta_{-17}}\biggr)_{\rm min} = \
\frac{28}{f_{\rm O_2}}
\ \ \ {\rm cm}^{-3}
\end{equation}
for which LIP solutions are possible.
For $f_{\rm O_2}=0.2$, this
is slightly smaller than the numerically computed
value of 370 cm$^{-3}$ at which the transition from the
LIP to HIP occurs (see Figure 1).
When the gas density falls below the critical density
electrons remove H$_3^+$, but O$_2$ continue to remove S$^+$.
The H$_3^+$-O$_2$-S$^+$ instability cycle therefore turns on,
and the gas is driven from the LIP to the HIP.

For consistency we must now verify that 
reactions with O$_2$
dominate the removal of S$^+$ everywhere in the LIP.
This requires that
\begin{equation}
\frac{k_{22}n({\rm O_2})}{k_{17}n({\rm e})} =
\frac{2X_{\rm O}^{\prime 2}}{X_{\rm S}}
\frac{k_{22}^2}{k_{17}p_{21}}
f_{\rm O_2}^2
\frac{n_{\rm H_2}}{\zeta} = 
5.1 f_{\rm O_2}^2 
\frac{n_{\rm H_2}}{\zeta_{-17}} \gg 1 \ \ \ .
\end{equation}
for all LIP densities, down to the minimal density.
Inserting $(n/\zeta_{-17})_{\rm min}$ as given by
equation~(\ref{Le6}) gives
$k_{22}n({\rm O_2})/k_{17}n({\rm e})=142 f_{\rm O_2}$.
We conclude that consistent LIP solutions are obtained even
for $f_{\rm O_2}$ as small as $\sim 0.01$ 
(for which the O$_2$ density in the LIP
is still much larger than in the HIP.) 

\section{Scaling}

The key result of the above analysis is that for our standard parameters
$(n/\zeta_{-17})_{\rm max}$, as given by equation~(\ref{Hcond2}), is
{\it greater} than $(n/\zeta_{-17})_{\rm min}$, as given by 
equation~(\ref{Le6}). The gas is therefore bistable
between the maximal HIP density and minimum LIP density.
The MIX branch must therefore ``switch back'' from the
maximal HIP density to the minimal LIP density. 
The H$_3^+$-O$_2$-S$^+$ instability cycle drives the O$_2$
density to either very low values in the HIP where a large
ionization fraction is maintained by slow radiative recombination,
or it drives the O$_2$ density to very large values in the LIP
where a low ionization fraction is maintained by rapid dissociative
recombination. Our analytic estimates for the 
two critical densities are in good
agreement with the numerical results displayed in Figure 1. 

Equations~(\ref{Hcond}) and (\ref{Le5}) show that the critical
densities depend on several parameters, including
the gas-phase elemental abundances of carbon, oxygen, sulfur and helium,
and on several specific chemical rate coefficients, including
$k_{3,4}$, the dissociative recombination rate coefficient of H$_3^+$.
Varying any of these parameters alters the range of densities
for which bistability occurs, or eliminates it entirely if
$(n/\zeta_{-17})_{\rm max}$ becomes smaller than $(n/\zeta_{-17})_{\rm min}$.

A dependence of bistability on $k_{3,4}$,
a rate coefficient that has been the subject of much investigation
and controversy (see e.g., McCall et al.~2004), was found in
numerical computations by Pineau des Forets \& Roueff (2000).
In their calculations (see their Fig.~4) they set
$n_{\rm H_2}/\zeta_{-17}=2.5\times 10^3$ cm$^{-3}$, and 
varied $k_{3,4}$ from $10^{-8}$ to $10^{-6}$ cm$^3$ s$^{-1}$.
They found that
the gas is in the LIP or HIP, for low or high values of
$k_{3,4}$, and that it is bistable for intermediate 
H$_3^+$ recombination efficiencies.
Our analysis provides an explanation for this behavior.

In Figure 4 we plot
$x_{\rm e}$ versus $n_{\rm H_2}/\zeta_{-17}$,
for five values of $k_{3,4}$, ranging from 0.2 to 5 times
our standard 50 K value of $1.7\times 10^{-7}$ cm$^3$ s$^{-1}$.
In these numerical computations, all other parameters, 
including all rate coefficients and gas-phase elemental abundances, 
are the same as in the computations we presented in \S 2.
Figure 4 shows that for small $k_{3,4}$ bistability is suppressed, and
that as $k_{3,4}$ is increased the gas becomes bistable for an
increasing range of densities.  For $n_{\rm H_2}/\zeta_{-17}$
held fixed at, say, $1.0\times 10^3$ cm$^{-3}$, the gas
is in the LIP or HIP for low or high $k_{3,4}$,
and is bistable for intermediate values. This is the same
behavior found by Pineau des Forets \& Roueff (2000).

These results can be readily understood as follows.
As the dissociative recombination rate coefficient of
H$_3^+$ is increased, less O$_2$ 
is produced by the sequence (R6)-(R11) at any give density.
An HIP can therefore be maintained to higher densities before
O$_2$ becomes significant in removing S$^+$.
An increased H$_3^+$ recombination efficiency also
implies that electrons become competitive with
CO in removing H$_3^+$ at larger densities in the LIP.
Quantitatively, equation~(\ref{Hcond}) shows that 
$(n_{\rm H_2}/\zeta_{-17})_{\rm max}$, the density above which
O$_2$ begins removing S$^+$, is proportional to $k_{3,4}^2$. However, 
$(n_{\rm H_2}/\zeta_{-17})_{\rm min}$, the density 
below which electrons begin removing the H$_3^+$,
varies only linearly with $k_{3,4}$. Therefore, as $k_{3,4}$ 
is reduced, $(n/\zeta_{-17})_{\rm max}$
falls {\it below} $(n_{\rm H_2}/\zeta_{-17})_{\rm min}$, 
and bistability disappears.
As $k_{3,4}$ is increased, $(n_{\rm H_2}/\zeta_{-17})_{\rm max}$ grows
more rapidly, and eventually exceeds,
$(n_{\rm H_2}/\zeta_{-17})_{\rm min}$, and the gas becomes bistable for
an increasing range of densities.
For example, if $k_{3,4}$ is increased by a factor of 5
relative to our standard value, 
Figure 4 shows that $(n_{\rm H_2}/\zeta_{-17})_{\rm max}$ increases
by about a factor of 25, and $(n_{\rm H_2}/\zeta_{-17})_{\rm min}$
increases by about a factor of 5. This is consistent with
expressions~(\ref{Hcond}) and (\ref{Le5}).

A similar scaling analysis may be carried out for the other
parameters that control $(n_{\rm H_2}/\zeta)_{\rm min}$
and $(n_{\rm H_2}/\zeta)_{\rm max}$. For example,
if the oxygen abundance is reduced our analysis shows that 
the regime of bistability should move to higher values
of $n_{\rm H_2}/\zeta$, as found numerically by 
Viti et al.~(2001) (see their Figures 5 and 7). 
For a lower oxygen abundance $(n_{\rm H_2}/\zeta)_{\rm min}$
is increased because
the HIP can be maintained to
a higher density before O$_2$ becomes abundant enough to
remove the atomic ions. The reduced O$_2$ density 
leads to a higher electron density in the LIP, so
$(n_{\rm H_2}/\zeta)_{\rm min}$ is increased as well.

Our numerical and analytic results show 
how the onset of the H$_3^+$-O$_2$-S$^+$ instability cycle
depends on the O$_2$ and electron densities as determined
by the elemental abundances and
by the H$_3^+$ dissociative recombination rate coefficient.
However, Figure 4 reveals an
additional interesting feature. For $k_{3,4}$ equal to 0.2
times our standard value the gas again becomes marginally
bistable at $(n_{\rm H_2}/\zeta_{-17})\approx 9$ cm$^{-3}$.
At such low densities the dominant atomic ion in the 
HIP is C$^+$ rather than S$^+$ (see Fig.~1). The instability,
and the transition from HIP to LIP,
is therefore modified by the addition of an
identical H$_3^+$-O$_2$-C$^+$ cycle, 
with reactions (R19) and (R20) operating to neutralize the C$^+$.
The behavior in this numerical example is a bit more complicated
because S$^+$ is still the dominant ion in the LIP at the
transition density.  We note that at such low densities
realistic clouds would likely become ``diffuse'' and optically
thin to external far-ultraviolet radiation, with a resulting
reduction in O$_2$ and further modifications to the chemistry. 
We do not consider such effects here.

\section{Gas-Grain Recombination}

We have shown that bistability arises in purely gas-phase models,
consistent with previous findings. 
When the HIP and LIP solutions ``overlap'', 
neutralization of atomic ions (S$^+$ in our models)
proceeds by either slow radiative recombination in the HIP,
or via the formation of intermediate molecular ions
(${\rm S^+ + O_2 \rightarrow SO^+ + O}$) followed by
rapid dissociative recombination in the LIP.
The two solutions are separated by a chemical 
instability, which we have identified
in our models as the H$_3^+$-O$_2$-S$^+$ cycle.

Our analysis suggests that bistability will be damped,
or eliminated entirely, if additional neutralization 
processes are included that interrupt the H$_3^+$-O$_2$-S$^+$ cycle.
We demonstrate this by considering how the single process of
``grain assisted recombination'' of atomic ions affects the chemistry.
In this process, atomic ions recombine via electron 
transfer from grains to the ions during collisions with the grains 
(Draine \& Sutin 1987; Weingartner \& Draine 2001).
The grain assisted recombination rate may be expressed as
$k_gn_{\rm tot}$, where $n_{\rm tot}$ is the total density
of hydrogen nuclei. For any ion, $k_g$ is an effective rate coefficient
that depends on the dust-to-gas ratio, grain size distribution,
grain charge, and gas temperature. 
This process dominates radiative recombination when $2k_g > k_rx_{\rm e}$,
where $k_r$ is the radiative recombination rate coefficient.  
For $k_r\sim 10^{-11}$ cm$^3$ s$^{-1}$ (appropriate for S$^+$), 
and a fractional ionization $x_{\rm e}\approx 10^{-5}$, as occurs for
densities near $(n_{\rm H_2}/\zeta_{-17})_{\rm max}$ in our computation 
(see Fig.~1), 
grain assisted recombination dominates radiative recombination
for $k_g\gtrsim 5\times 10^{-17}$ cm$^3$ s$^{-1}$. This is well
below the maximum values of $k_g \sim 10^{-14}$ cm$^3$ s$^{-1}$
found by Weingartner \& Draine, so this process is potentially
relevant in determining the ionization balance.
Furthermore, grain assisted recombination will
interfere with the H$_3^+$-O$_2$-S$^+$ cycle and render it inoperative
when $k_g > f_{\rm O_2}X^\prime_{\rm O}k_{22}$, where as previously
defined, $f_{\rm O_2}X^\prime_{\rm O}$ is the fraction of oxygen 
bound in O$_2$, and $k_{22}=1.5\times 10^{-11}$ cm$^3$ s$^{-1}$
is the rate coefficient of reaction (R22). For $f_{\rm O_2}\approx0.2$ and 
$X^\prime_{\rm O}=1.5\times 10^{-4}$, the H$_3^+$-O$_2$-S$^+$ instability
cycle is quenched when $k_d\gtrsim 5\times 10^{-16}$ cm$^3$ s$^{-1}$.
Bistability and the sharp transition from HIP to LIP should then disappear.

We illustrate this in Figure 5, where
we plot $x_{\rm e}$ versus $n_{\rm H_2}/\zeta_{-17}$, for $k_g$ ranging
from zero (as in Figures 1 and 4) to $1.0\times 10^{-14}$ cm$^3$ s$^{-1}$,
for all atomic ions. All other parameters are as in the calculation
in \S 2.
As $k_g$ is increased the diminished electron density leads to
more H$_3^+$ and O$_2$ in the HIP, so that 
$(n_{\rm H_2}/\zeta_{-17})_{\rm max}$ is reduced. Similarly,
the onset of dissociative recombination in the LIP occurs at
lower densities, and $(n_{\rm H_2}/\zeta_{-17})_{\rm min}$ is reduced.
Thus, at first the bistability region shifts
to lower densities as $k_g$ is increased. However, as estimated above, when 
$k_g$ exceeds $\sim 3\times 10^{-16}$ cm$^3$ s$^{-1}$ bistability
is eliminated entirely, because the H$_3^+$-O$_2$-S$^+$ (and
H$_3^+$-O$_2$-C$^+$) cycles are fully quenched, and neutralization
occurs entirely by grain recombination.

These results strongly suggest that for realistic clouds, 
in which gas-grain neutralization is a likely important process, 
bistability will not occur.
The precise value of $k_g$ in dark clouds may be variable, as the
grain population can be modified by coagulation 
(Draine 1985; Flower et al.~2005) and other processes. However,
even with the removal of
all grains with sizes below $\sim 15$ \AA, the rate coefficient
$k_g\sim 10^{-15}$ cm$^3$ s$^{-1}$ (e.g.~McKee 1989), which is sufficiently
large to suppress bistability.
We thus confirm and explain
the results of Shalabiea and Greenberg (1995),
who found that bistability is eliminated
when gas-grain processes, including grain assisted recombination,
are incorporated in the numerical computations. 

\section{Summary} 

The non-linearity of equations (1)-(4) is a mathematical
prerequisite for bistability and the appearance of multiple solutions
(Le Bourlot et al.~1993, 1995; Lee et al.~1998; 
Pineau des Forets \& Roueff 2000; Charnley \& Markwick 2003). 
However, the purpose of our paper has been to identify and
analyze the chemical mechanisms 
underlying the bistability phenomenon. We have shown that
a simple chemical instability, involving atomic ions (typically S$^+$),
O$_2$ molecules, and H$_3^+$ ions, allows multiple solutions
to appear near the transition points from low-density high-ionization
(HIP) to high-density low-ionization (LIP) conditions in the
gas-phase chemistry.  The instability drives the O$_2$ density to
either small (HIP) or large (LIP) values. Bistability occurs
when a low O$_2$ abundance in the HIP can be maintained up to
maximal densities that are greater than the minimal densities
above which a large O$_2$ density is maintained in the LIP.
Two modes of
neutralization are then simultaneously available to the gas.
The slow (HIP) mode is radiative recombination
(usually ${\rm S^+ + e \rightarrow S + \nu}$).
The fast (LIP) mode is formation of molecular ions
followed by dissociative recombination
(usually ${\rm S^+ + O_2 \rightarrow SO^+ + O}$;
${\rm SO^+ + e \rightarrow S + O}$).

We have derived simple analytic scaling relations for the
range of gas densities and ionization rates for which
bistability occurs. Our formulae show how the bistable range
depends on the assumed gas-phase abundances, and
several reaction rate-coefficients, including the
dissociative recombination rate coefficient of H$_3^+$.
Bistability is eliminated when the two modes of gas-phase
neutralization are superseded by gas-grain processes
such as grain assisted recombination of 
the atomic ions. This likely happens for the range of
electron and O$_2$ densities where bistability 
typically occurs in purely gas-phase models.
We conclude that the bistability phenomenon probably does not
occur in realistic dusty interstellar clouds, although it
remains of theoretical interest.

     
\section*{Acknowledgments}

We thank A.~Dalgarno, O.~Gnat, C.F.~McKee, and D.~Neufeld for
discussions. We thank A.~Markwick-Kemper for constructive
comments on our manuscript.
This research is supported by the Israel 
Science Foundation, grant 221/03.


\clearpage

\begin{figure}
\plotone{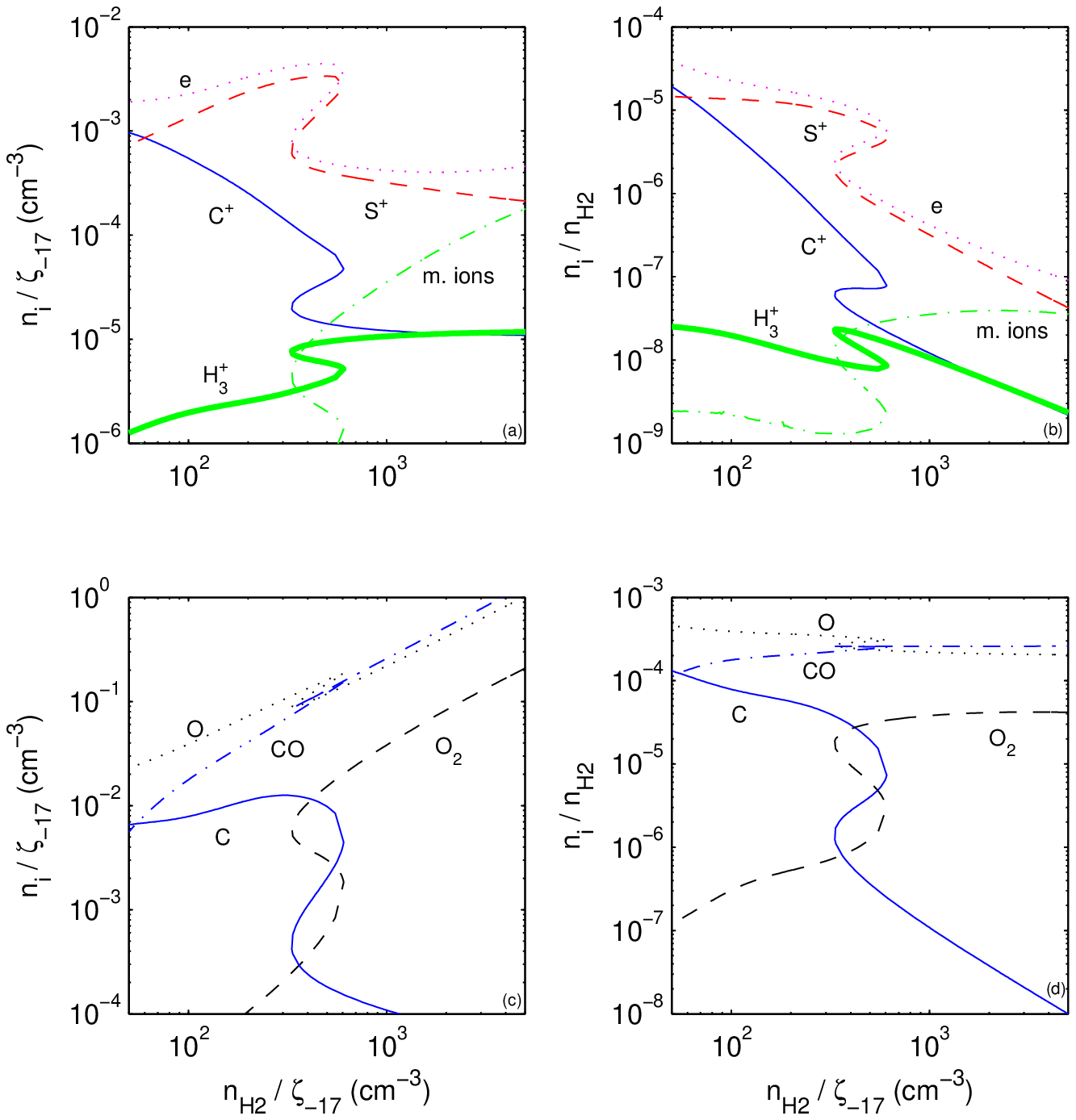}
\caption{The steady-state densities $n_i/\zeta_{-17}$, and density fractions
$n_i/n_{\rm H_2}$, of electrons, S$^+$, C$^+$, and H$_3^+$ ions,
C, and O atoms, and CO and O$_2$ molecules, as functions of 
$n_{\rm H_2}/\zeta_{-17}$. An HIP exists for densities
less than 675 cm$^{-3}$, and a LIP exists for densities greater
than 370 cm$^{-3}$. The gas is bistable between these densities.
  } 
\label{prates}
\end{figure}

\clearpage

\begin{figure}
\includegraphics[width=13.17cm,height=19cm]{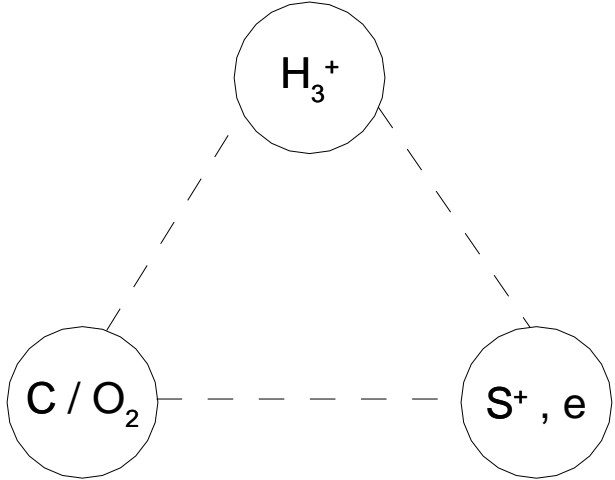}
\caption{The H$_3^+$--O$_2$--S$^+$ cycle.
Increasing the electron density, reduces H$_3^+$ via dissociative
recombination. This reduces O$_2$ formed by the sequence
(R6)-(R11). Destruction of S$^+$ by  O$_2$ decreases,
so the S$^+$ density increases.
This cycle can ``run away'' when S$^+$ is the dominant
positive charge carrier (see text).
} 
\label{prates}
\end{figure}

\clearpage

\begin{figure}
\plotone{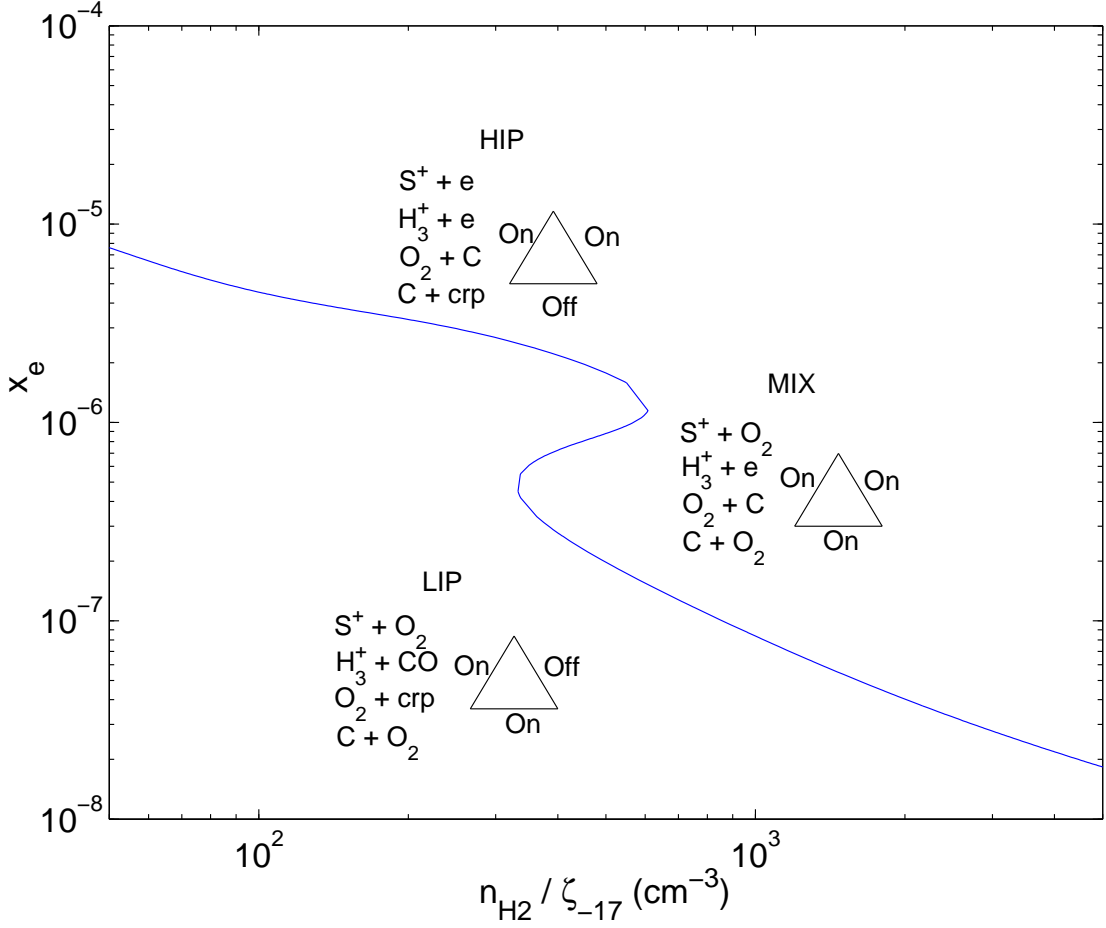}
\caption{The fractional ionization, $x_{\rm e}\equiv n({\rm e})/n_{\rm H_2}$,
 versus $n_{\rm H_2}/\zeta_{-17}$,
for our standard model. The dominant destruction reactions
for S$^+$, H$_3^+$, O$_2$, and C, are indicated for each
of the three HIP, MIX, and LIP branches in the bistability region.
E.g., ``{\rm S$^+$~+~e}'' indicates that S$^+$ is removed by recombination
with electrons. The triangles indicate which ``legs'' of the
H$_3^+$--O$_2$--S$^+$ instability cycle operate in each of the
three phases. 
} 
\label{prates}
\end{figure}

\clearpage

\begin{figure}
\plotone{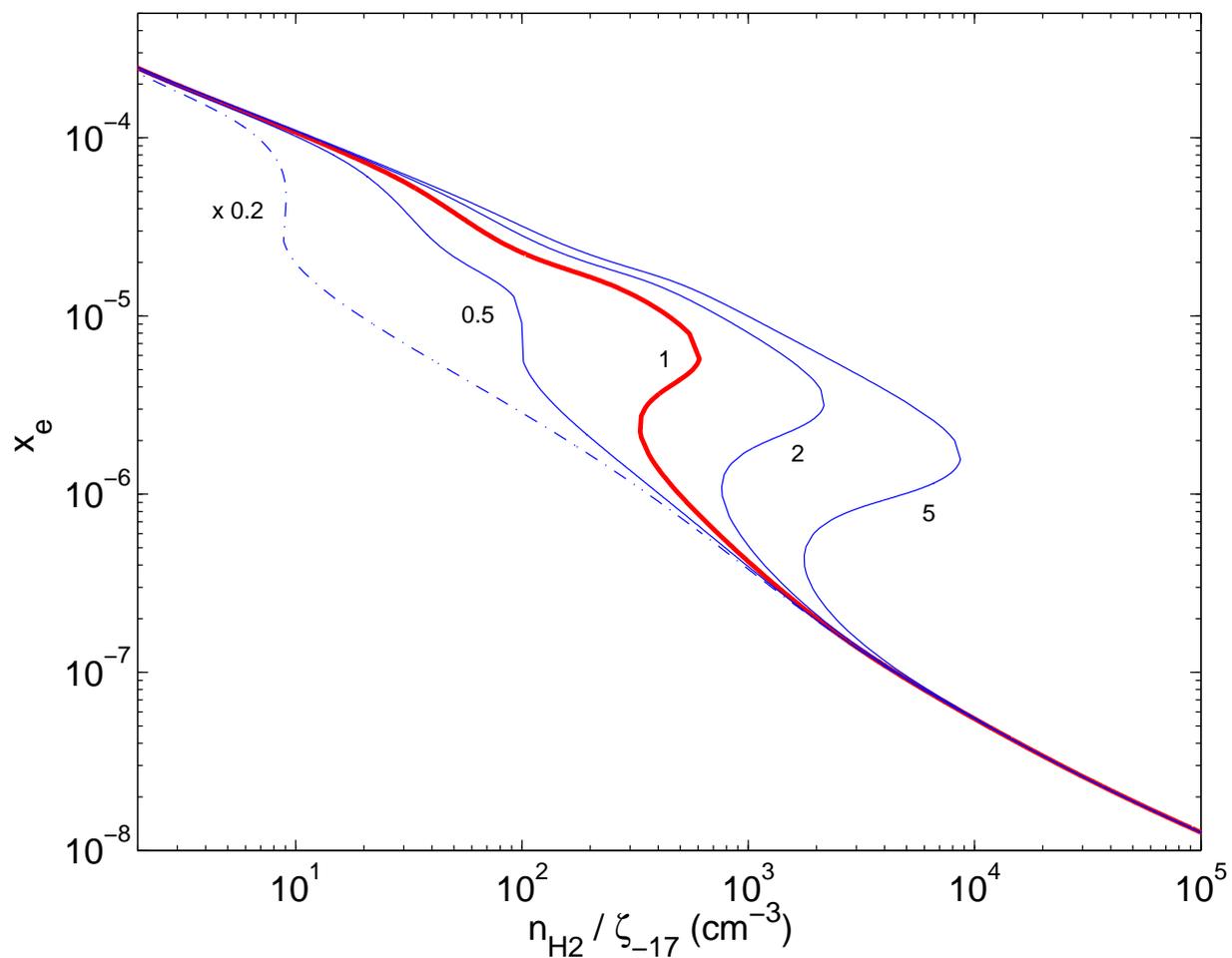}
\caption{The fractional ionization, $x_{\rm e}\equiv n_{\rm e}/n_{\rm H_2}$,
 for five values of
the H$_3^+$ dissociative recombination rate coefficient $k_{3,4}$,
ranging from 0.2 to 5 times our standard value of 
$1.7\times 10^{-7}$ cm$^3$ s$^{-1}$. For the lowest value of
$k_{3,4}$ (dot-dashed curve) C$^+$ rather than S$^+$ is the dominant positive
ion near the transition point in the HIP (see text).
} 
\label{prates}
\end{figure}

\clearpage

\begin{figure}
\plotone{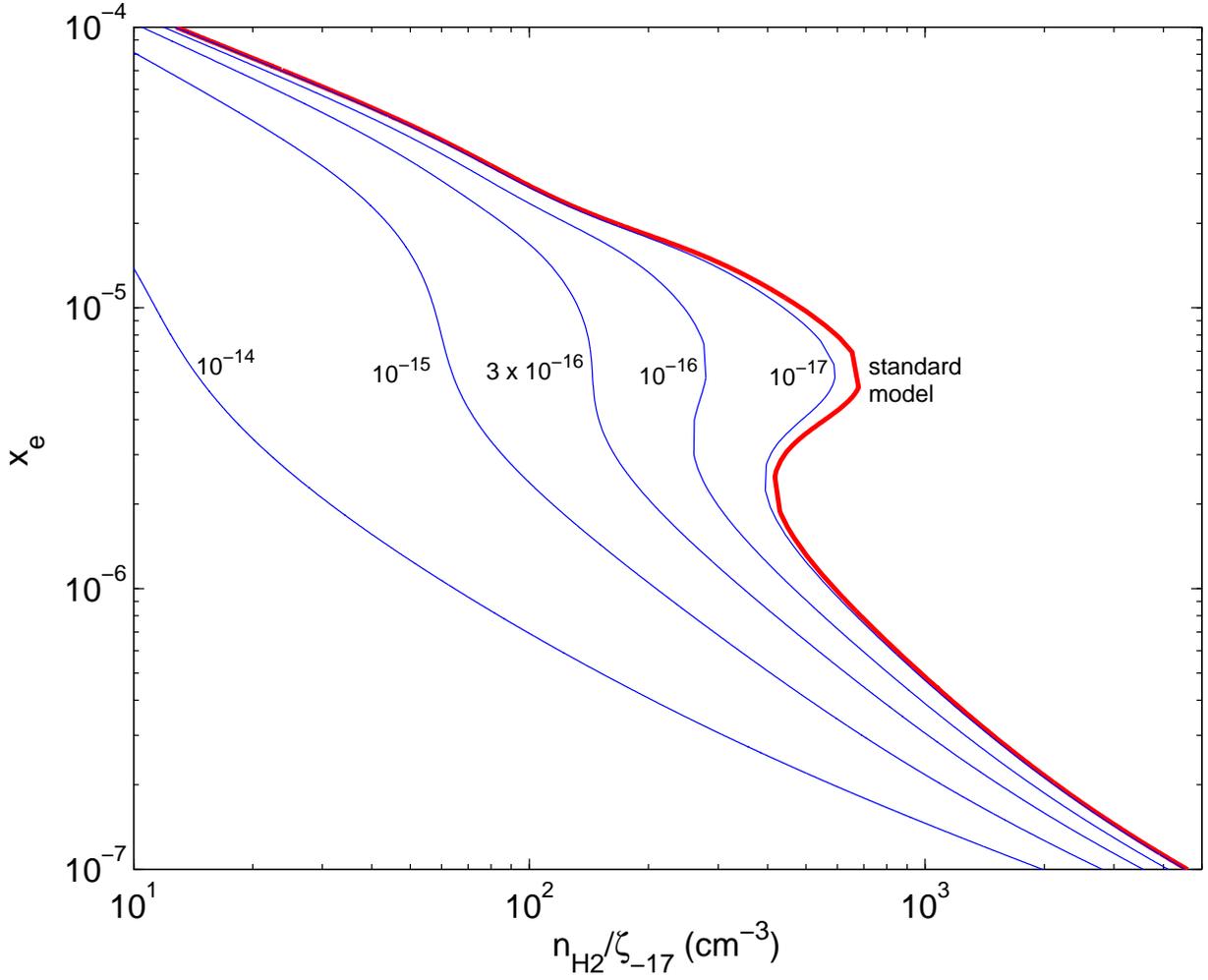}
\caption{The fractional ionization for
grain-assisted recombination rate-coefficients
for all atomic ions varying from $k_g=1.0\times 10^{-17}$ to
$1.0\times 10^{-14}$ cm$^3$ s$^{-1}$. For our
standard model, $k_g=0$.
} 
\label{prates}
\end{figure}

\clearpage
\begin{deluxetable}{lllllllll}
\footnotesize
\tablecaption{Atomic and molecular densities, and density ratios.}
\tablehead{
 & & HIP & & MIX & & LIP & & ratio
}
\startdata
H & & 1.7(-1) & & 1.7(-1) & & 1.7(-1) & & 1.0 \\
H$_2$ & & 4.3(2) & & 4.3(2) & & 4.3(2) & & 1.0 \\
C & & 1.1(-2) & & 1.3(-3) & & 2.2(-4) & & 50.0 \\
N & & 2.8(-2) & & 1.9(-2) & & 1.0(-2) & & 2.8 \\
O & & 1.4(-1) & & 1.2(-1) & & 1.0(-1) & & 1.4 \\
S & & 2.6(-3) & & 3.8(-3) & & 3.6(-3) & & 0.72 \\
Si & & 2.1(-5) & & 2.9(-5) & & 2.7(-5) & & 0.78 \\
e & & 4.4(-3) & & 1.7(-3) & & 5.3(-4) & & 8.30 \\
C$^+$ & & 9.0(-5) & & 3.1(-5) & & 1.5(-5) & & 6.0 \\
N$^+$ & & 5.0(-10) & & 5.4(-10) & & 6.0(-10) & & 0.83 \\
O$^+$ & & 4.2(-11) & & 1.1(-10) & & 1.9(-10) & & 0.22 \\
S$^+$ & & 3.3(-3) & & 1.3(-3) & & 4.3(-4) & & 7.67 \\
Si$^+$ & & 9.8(-4) & & 2.7(-4) & & 6.1(-5) & & 16.1\\
H$_3^+$ & & 3.6(-6) & & 6.3(-6) & & 9.0(-6) & & 0.40 \\
H$_3$O$^+$ & & 9.6(-8) & & 4.4(-7) & & 1.9(-6) & & 0.051 \\
HCO$^+$ & & 2.3(-7) & & 1.2(-6) & & 5.2(-6) & & 0.044 \\
HCS$^+$ & & 7.3(-8) & & 2.1(-7) & & 9.6(-7) & & 0.076 \\
CO$^+$ & & 2.2(-11) & & 5.8(-11) & & 9.5(-11) & & 0.23 \\
NO$^+$ & & 1.1(-8) & & 4.6(-8) & & 1.2(-7) & & 0.092 \\
O$_2^+$ & &  2.5(-9) & & 2.8(-8) & & 1.9(-7) & & 0.013 \\
SO$^+$ & & 3.7(-8) & & 1.6(-7) & & 5.2(-7) & & 0.071 \\
CH & & 2.7(-6) & & 7.3(-7) & & 3.0(-7) & & 9.0 \\
NH & & 1.1(-6) & & 1.5(-6) & & 2.0(-6) & & 0.55 \\
OH & & 2.8(-5) & & 6.3(-5) & & 1.1(-4) & & 0.25 \\
H$_2$O & & 2.8(-4) & & 1.1(-3) & & 2.1(-3) & & 0.13 \\
SH & & 3.9(-7) & & 6.1(-7) & & 1.0(-6) & & 0.39 \\
SiH & & 1.6(-7) & & 5.0(-8) & & 1.2(-8) & & 13.3 \\
CO & & 1.0(-1) & & 1.1(-1) & & 1.1(-1) & & 0.91\\
NO & & 3.0(-5) & & 1.2(-4) & & 3.0(-4) & & 0.1 \\
O$_2$ & & 4.2(-4) & & 3.6(-3) & & 1.2(-2) & & 0.035 \\
SO & & 6.5(-6) & & 1.2(-4) & & 8.0(-4) & & 0.008\\
SiO & &  5.3(-4) & & 1.2(-3) & & 1.4(-3) & & 0.38 \\
CS & & 1.2(-3) & & 1.8(-3) & & 2.2(-3) & & 0.55 \\
CN & & 4.5(-6) & & 2.0(-6) & & 1.3(-6) & & 3.46 \\
HCN & & 4.6(-5) & & 1.0(-5) & & 3.2(-6) & & 14.4 \\ 
N$_2$ & & 1.8(-2) & & 2.2(-2) & & 2.7(-2) & & 0.67\\
\enddata
\tablecomments{Densities (cm$^{-3}$) of selected
species, for the three HIP, MIX, and LIP solutions,
for $n_{\rm H_2}=4.3\times 10^2$ cm$^{-3}$, and
$\zeta=1.0\times 10^{-17}$ s$^{-1}$. Column 5 lists
HIP/LIP density ratios.}

\end{deluxetable}


\end{document}